\documentclass[conference]{IEEEtran}
\IEEEoverridecommandlockouts
\usepackage{cite}
\usepackage{amsmath,amssymb,amsfonts}
\usepackage{algorithmic}
\usepackage{graphicx}
\usepackage{textcomp}
\usepackage{xcolor}
\usepackage{multirow}
\usepackage[super]{nth}
\usepackage{hyperref}
\usepackage{lipsum}
\usepackage{dsfont}
\usepackage{booktabs}
\def\BibTeX{{\rm B\kern-.05em{\sc i\kern-.025em b}\kern-.08em
    T\kern-.1667em\lower.7ex\hbox{E}\kern-.125emX}}
\begin{document}
\title{Pioneering EEG Motor Imagery Classification \\Through Counterfactual Analysis\\
\thanks{This work was supported by the Institute of Information \& Communications Technology Planning \& Evaluation (IITP) grant, funded by the Korea government (MSIT) (No. 2019-0-00079, Artificial Intelligence Graduate School Program (Korea University)) and was partly supported by the Institute of Information \& communications Technology Planning \& Evaluation (IITP) grant, funded by the Korea government (MSIT) (No. 2021-0-02068, Artificial Intelligence Innovation Hub).}
}

\author{
\IEEEauthorblockN{Kang Yin}
\IEEEauthorblockA{\textit{Dept. of Artificial Intelligence} \\
\textit{Korea University}\\
Seoul, Republic of Korea \\
charles\_kang@korea.ac.kr}\\
\IEEEauthorblockN{Hee-Dong Kim}
\IEEEauthorblockA{\textit{Dept. of Artificial Intelligence} \\
\textit{Korea University}\\
Seoul, Republic of Korea \\
hd\_kim@korea.ac.kr}
\and
\IEEEauthorblockN{Hye-Bin Shin}
\IEEEauthorblockA{\textit{Dept. of Brain and Cognitive Engineering} \\
\textit{Korea University}\\
Seoul, Republic of Korea \\
hb\_shin@korea.ac.kr}\\
\IEEEauthorblockN{Seong-Whan Lee}
\IEEEauthorblockA{\textit{Dept. of Artificial Intelligence} \\
\textit{Korea University}\\
Seoul, Republic of Korea \\
sw.lee@korea.ac.kr}
}

\maketitle
\begin{abstract}
The application of counterfactual explanation (CE) techniques in the realm of electroencephalography (EEG) classification has been relatively infrequent in contemporary research. In this study, we attempt to introduce and explore a novel non-generative approach to CE, specifically tailored for the analysis of EEG signals. This innovative approach assesses the model's decision-making process by strategically swapping patches derived from time-frequency analyses. By meticulously examining the variations and nuances introduced in the classification outcomes through this method, we aim to derive insights that can enhance interpretability. The empirical results obtained from our experimental investigations serve not only to validate the efficacy of our proposed approach but also to reinforce human confidence in the model's predictive capabilities. Consequently, these findings underscore the significance and potential value of conducting further, more extensive research in this promising direction. The implementation is available at \href{https://github.com/Kang1121/EEG-Counterfactual}{https://github.com/Kang1121/EEG-Counterfactual}.
\end{abstract}

\begin{small}
\end{small}

\section{INTRODUCTION}

The evolution of electroencephalograph (EEG) analysis and brain-computer interfaces (BCI) has greatly benefited from emerging computational techniques. Kim \textit{et al.}\cite{b1} highlight the integration of convolutional neural network (CNN) in single-trial electromyography (EMG) analysis. Thung \textit{et al.}\cite{b3} delve into predictive modeling for cognitive impairments using matrix completion techniques. Ma \textit{et al.}\cite{add1} integrate CNN with attention for motor imagery (MI) classification. In parallel, Lee \textit{et al.}\cite{b5} have showcased the potential influence of auditory stimuli on brain functions, and Roy~\cite{add2} fuses multi-scale features for multi-classification. Such advancements coexist with intriguing research into the neural underpinnings of emotion representation\cite{b9} and state-of-the-art multi-view CNN architectures\cite{b10} for BCI.

But while these methods show promise, it's unclear how they make predictions about EEG data. The comprehension of deep learning, indeed, can often be challenging and complex. Some research\cite{b11,b12} try to explain EEG models using tools like shapley additive explanations (SHAP)\cite{b13} and layer-wise relevance propagation (LRP)\cite{b14}, focusing on the importance of certain features or neurons. Yet, these tools may not offer clear or actionable insights. Another approach, counterfactual explanation (CE), offers more direct advice by suggesting specific changes to input data. While CE is popular in image studies, it's less common in brain-machine research. Our paper looks at how CE might help clarify EEG model predictions, drawing from previous research\cite{b15} in image analysis.

In this study, we apply CE to vividly illustrate how a model gives its predictions from specific regions within EEG spectrogram. We emphasize the connection between the counterfactuals and their original data samples. Our findings indicate that implementing CE in EEG studies is not only feasible but also holds considerable potential for further exploration. This research makes two significant advancements:
\begin{itemize}
\item We present an innovative approach by employing CE to shed light on EEG classifications. As far as we're aware, ours is a pioneering effort in this domain with EEG signals.
\item We consciously bypass generative CE. Such a decision guarantees that the resulting CE remains consistent with genuine, credible data and is particularly apt for the more compact datasets common in EEG research.
\end{itemize}

\section{METHODOLOGY}
\subsection{Problem Formulation}
Consider an EEG spectrogram dataset $\mathcal{I}$ with $\mathcal{C}$ classes. For a class $c$ with $n$ samples, let query set $Q=\{I_i^{c}\}_{i=1}^n$ and distractor set $D=\bigcup_{j=1,j\neq c}^{\mathcal{C}}\{I_i^{j}\}_{i=1}^n$, such that $Q\cup D=\mathcal{I}$ and $Q\cap D = \varnothing$. Let $I$ and $I'$ be random samples from $Q$ and $D$ with classes $c$ and $c'$.

Building upon the work in\cite{b15}, consider a feature extractor $f:\mathcal{I}\to \mathbb{R}^{hw\times d}$ that transforms the EEG spectrogram into a feature dimension of $h\times w\times d$. Alongside, there's a classifier $g:\mathbb{R}^{hw\times d}\to\mathbb{R}^{\mathcal{C}}$ that associates the spatial feature to a set of $\mathcal{C}$-logits.  We formulate a counterfactual $I^*$ within the feature space $f(\cdot)$ by substituting spatial cells in $f(I)$ with those from $f(I')$, ensuring that the classifier outputs $c'$ for $I^*$. The counterfactual $I^*$ is expressed as:
\begin{equation}
    f(I^*)=(\mathds{1} -\mathbf{a})\circ f(I) + \mathbf{a}\circ Pf(I'),
\end{equation}
where $\mathds{1}$ stands for a vector filled with ones. The sparse binary gating vector $\mathbf{a}\in \mathbb{R}^{hw}$ adjusts the mix of $f(I')$ and $f(I)$ in the resultant $f(I^*)$. $P\in \mathbb{R}^{hw\times hw}$ is a permutation matrix ensuring the alignment of $f(I')$ and $f(I)$, and $\circ$ denotes the Hadamard product.

\subsection{Objective Function}
Our objective is to alter the model's prediction from $c$ to $c'$. Specifically, we aim to maximize the probability of the modified feature $f(I^*)$  being associated with class $c'$. The objective is defined as:
\begin{equation}
    \max_{P,a} g_{c'}((\mathds{1} -\mathbf{a})\circ f(I) + \mathbf{a}\circ Pf(I')),
    \label{loss}
\end{equation}
where $P\in \mathcal{P}$, the collection of all $hw \times hw$ permutation matrices. Drawing from the concepts in CE\cite{b15,b16}, our interest lies in identifying the minimal number of edit operations needed for a prediction shift. As such, to avoid learning an identical mapping, Eq.~\ref{loss} incorporates regularization. This is characterized by the constraints $||\mathbf{a}||_1=1, \mathbf{a}_i\in\{0, 1\}$. Given that $\mathbf{a}$ is binary, reducing its norm equates to minimizing the edit count between $I'$ to $I$.

\section{EXPERIMENT}
\subsection{Dataset and Preprocessing}
In this research, we utilize the BCI-C IV 2a EEG dataset\cite{b17}. This dataset comprises EEG recordings from two sessions involving 9 healthy participants, each performing 4 MI tasks. Each task consists of 72 trials per session. The EEG recordings were captured through 22 channels at a sampling rate of 250 Hz, with a 3-second MI period for every trial. In this study, we only choose left-and-right-hand MI tasks in the following experiments.

To adapt the data for image processing suitable for network input\cite{b16}, we first downsample the signal by a factor of 3. Subsequently, we clip the data and select 224 linearly spaced frequencies from 1 Hz to 41 Hz. We then employ the continuous wavelet transform (CWT) to obtain the spectrogram. It is important to note that the use of the EEG spectrogram as an input to the model is a well-founded choice, substantiated by preceding research \cite{b6, b4, spectro, b2}, which has demonstrated that spatio-temporal attributes frequently yield robust performance metrics. Once the preprocessing phase is complete, the EEG data is reshaped into a 224 $\times$ 224 matrix, with the first axis representing the frequency dimension and the second representing the temporal dimension. 

\subsection{Experimental Setup and Classification Results}
Table~\ref{dataset_overview} displays our approach to data categorization. While traditionally, left-hand MI and right-hand MI are treated as two separate classes, we introduce another configuration where each subject is considered its own domain. This results in 2 $\times$ 9 = 18 classes, leading to a reduction in sample size by a factor of 9. Besides, our training methodology diverges from typical practices. We allocate half of each subject's data for training, split the remaining half between validation and test for CE. The ResNet-18 framework, embodying functions $f$ and $g$, serves as our primary model. Instead of utilizing pretrained models, we train from scratch, leveraging MixUp to enhance the training process. The Top-1 accuracy for 2-class and 18-class settings are 69.44 \% and 54.78 \%, respectively.

\begin{table}[!tb]
    \centering
    \caption{Overview of experimental setup and Top-1 accuracy.}
    \begin{tabular}{llllc}
        \toprule
        \textbf{Dataset} & \multicolumn{3}{c}{\textbf{Statistics}} & \textbf{Top-1 (\%)}\\
        \cmidrule(lr){2-4} \cmidrule(lr){5-5}
        & \# Class & \# Train & \# Val & ResNet-18\\
        \midrule
        BCI-C IV 2a (2-class) & 2 & 1,296 & 648 & 69.44\\ 
        BCI-C IV 2a (18-class) & 18 &1,296&  648& 54.78\\  
        \bottomrule
    \end{tabular}
    \label{dataset_overview}
\end{table}
\begin{figure}[tb]
    \centering
    \begin{minipage}{0.45\linewidth}
        \includegraphics[width=\linewidth]{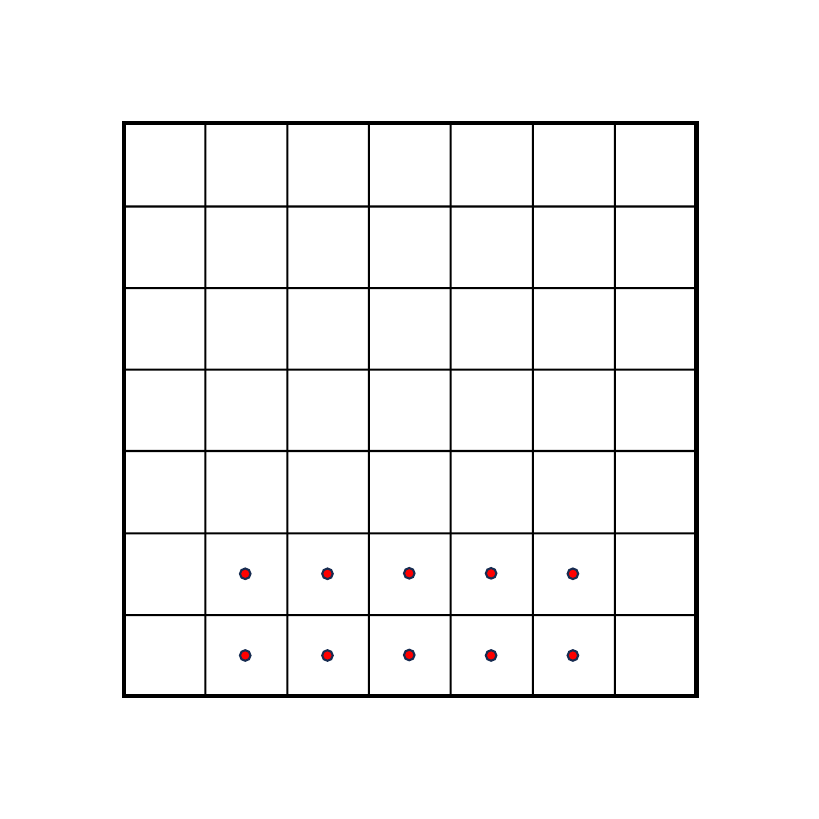}
    \end{minipage}
    \begin{minipage}{0.45\linewidth}
        \includegraphics[width=\linewidth]{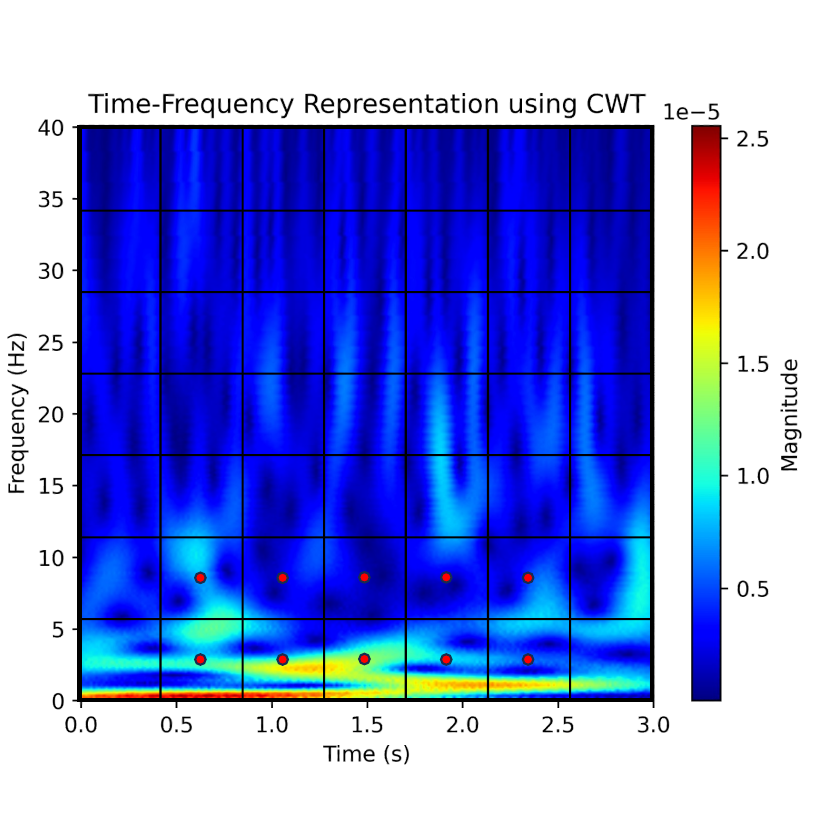}
    \end{minipage}
    \caption{Configuration of (left) keypoint annotations and (right) the corresponding areas in the spectrogram.}
    \label{anno}
\end{figure}
\subsection{Evaluation Metrics}
We adopt the evaluation metrics as outlined in \cite{b16}:
\begin{itemize}
    \item \textbf{Near-KP} assesses whether the spectrogram regions contain object keypoints.
    \item \textbf{Same-KP} quantifies the frequency with which we select identical keypoints in both the query and distractor spectrograms.
    \item \textbf{\# Edits} calculates the average count of modifications required for the classification model to predict the distractor class $c'$ on the altered spectrogram $I^*$.
\end{itemize}

A crucial metric in vision CE is the keypoint (KP) evaluation, which relies on manually annotated object coordinates within images. For our specific task, we establish keypoints as detailed in Fig.~\ref{anno} and assess our results based on these keypoints. We select a total of 10 keypoints, centrally located in the middle regions of the final two rows. Our rationale for choosing these particular points is twofold. From a frequency perspective, they encompass the entirety of the $\alpha$ band, which has been demonstrated to play a significant role in MI tasks. Additionally, they touch upon the $\beta$, $\theta$, and $\delta$ bands, all of which may be involved in the MI task to some extent. Temporally speaking, under the assumption that the data follows a normal distribution, we focus on the central portion, which is where MI activity is most likely to occur.

\subsection{Quantitative Analysis}
\begin{table}[tb]
    \centering
    \caption{Counterfactual results on BCI-C IV 2a dataset.}
    \begin{tabular}{llccc}
        \toprule
        \textbf{Setting} &&\multicolumn{3}{c}{\textbf{BCI-C IV 2a}}\\
        \cmidrule{3-5}
        & & Near-KP (\%)& Same-KP (\%)& \# Edits \\
        \midrule
        Single & 2-class & 92.88 & 19.55& -\\
        Edit & 18-class & 74.50& 13.80&  -\\
        \midrule
        All & 2-class & 81.80& 14.51& 2.08\\
        Edits & 18-class & 71.15& 11.19& 2.49\\
        \bottomrule
    \end{tabular}

    \label{results}
\end{table}
Table~\ref{results} provides results in two distinct settings: a "Single Edit" scenario and an "All Edits" scenario. The former represents a one-time solution to Eq.~\ref{loss}, while the latter involves iteratively solving the same equation until a shift in prediction is observed. A notable observation is the elevated scores for Near-KP. This signifies alterations primarily at the annotated keypoint regions, aligning with empirical evidence. Concurrently, we note an exceptionally low score for Same-KP. This may imply that the model discerns distinct, non-overlapping patterns to differentiate between classes. Additionally, for the 2-class and 18-class categorizations, the average edit times stand at 2.08 and 2.49 respectively. This suggests the model's capability in identifying distinctive features, enabling it to differentiate between class variations.

\subsection{Qualitative Results}

\begin{figure}[tb]
    \centering
    \resizebox{0.9\linewidth}{!}{
    \includegraphics{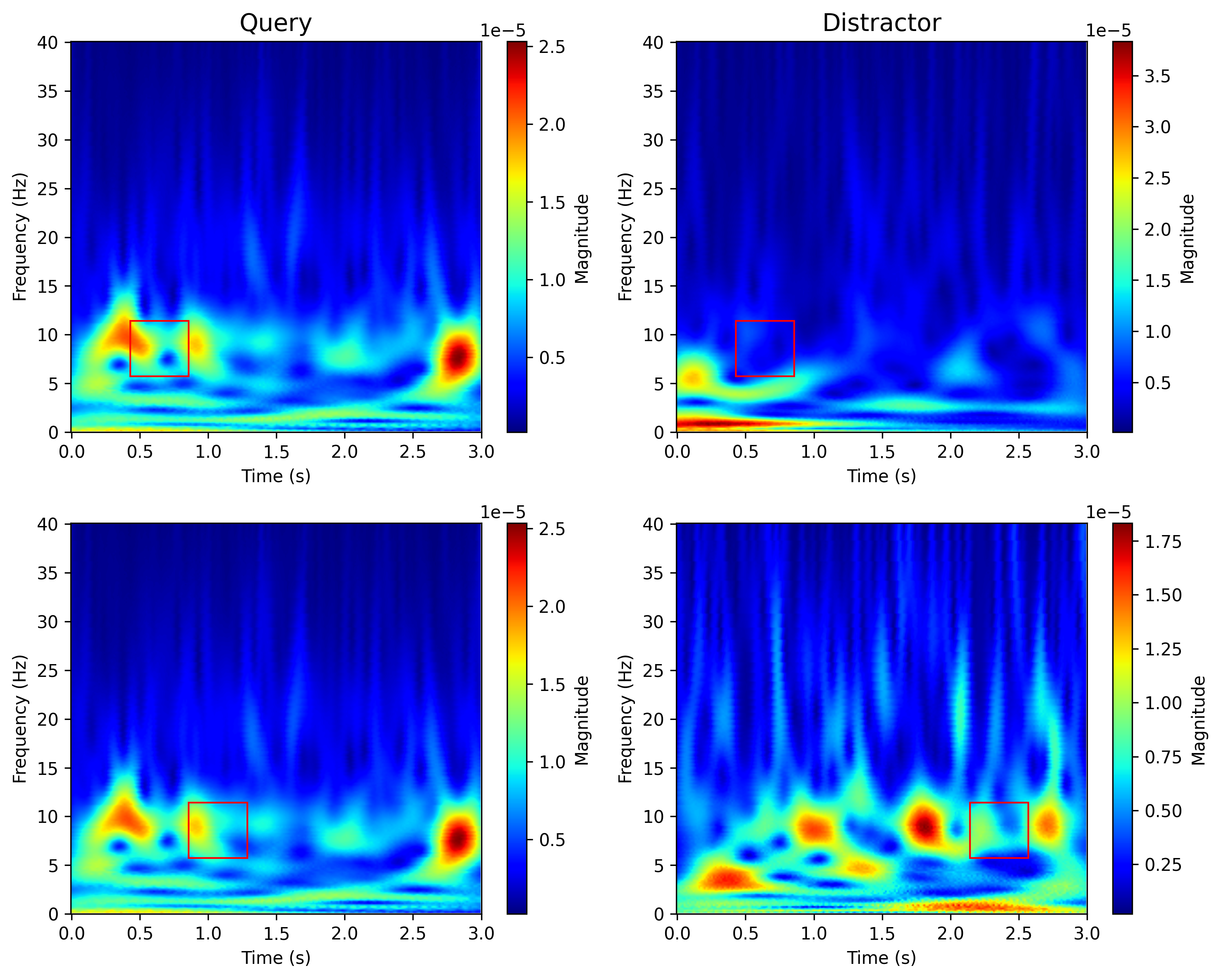}}
    \caption{Visualization of the replacing regions in query and distractor spectrograms. (\# Edits = 2)}
    \label{vis_spec}
\end{figure}

\begin{figure*}[tb]
    \centering    
    \begin{minipage}{0.75\textwidth}
        \centering
        \includegraphics[width=0.32\linewidth]{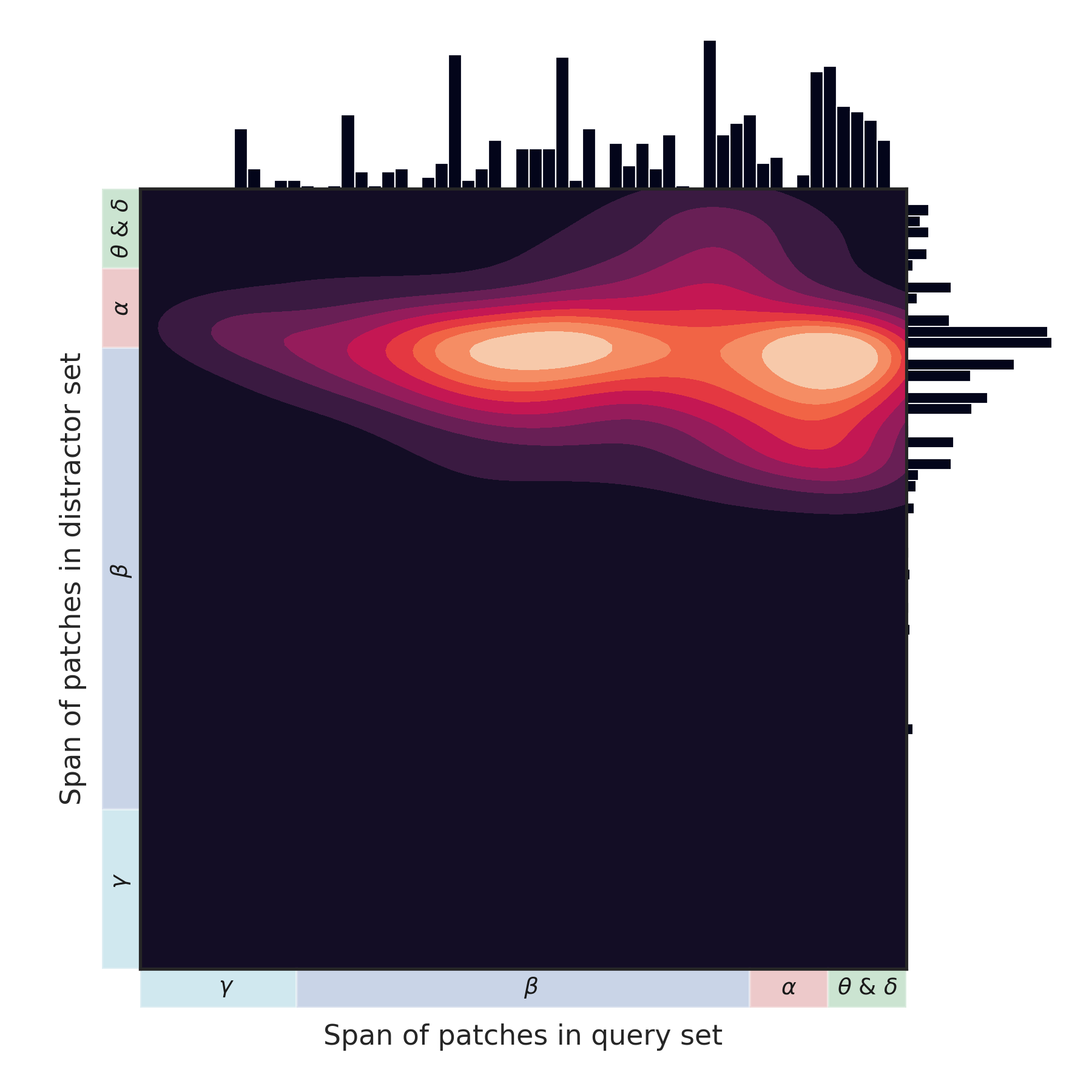}
        \includegraphics[width=0.32\linewidth]{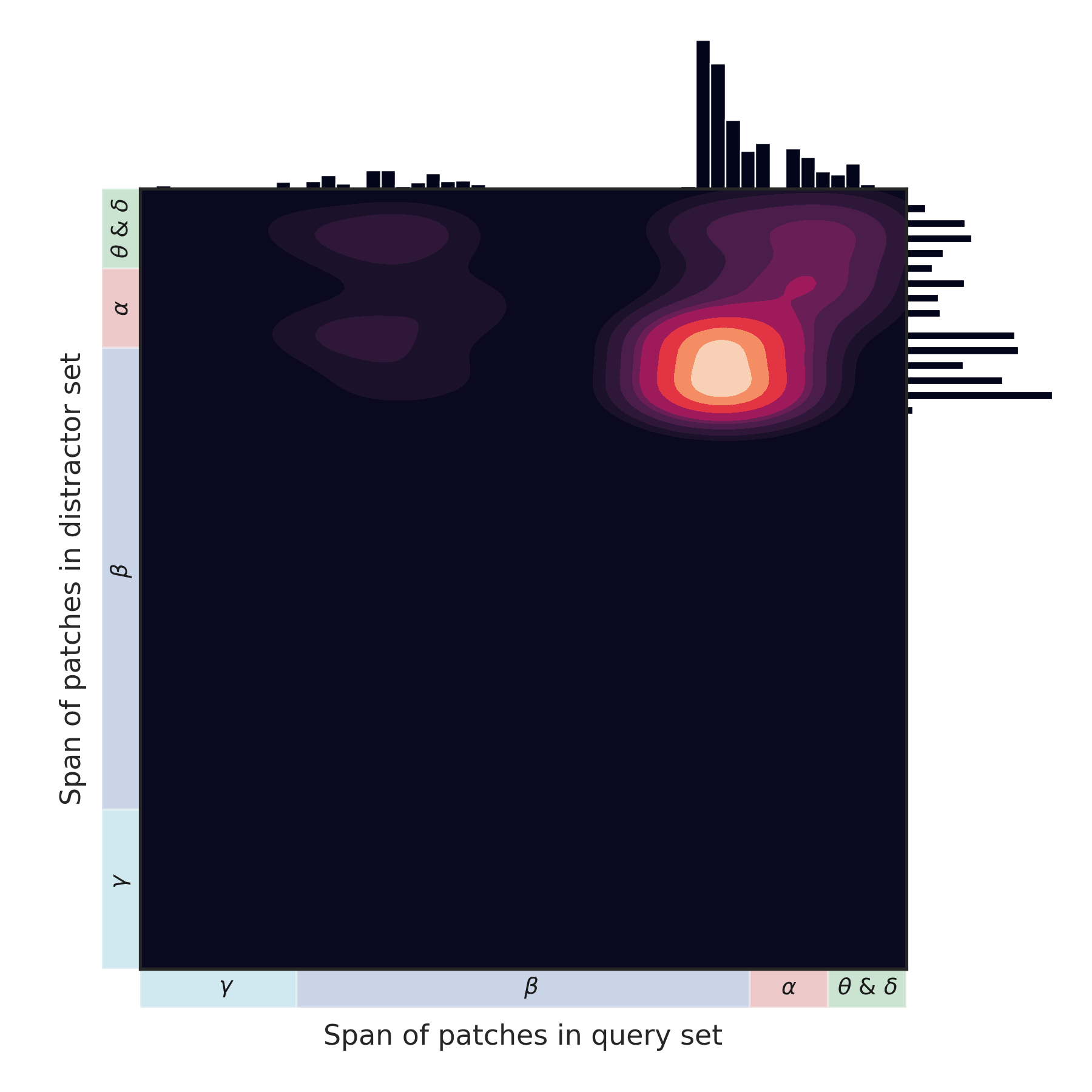}
        \includegraphics[width=0.32\linewidth]{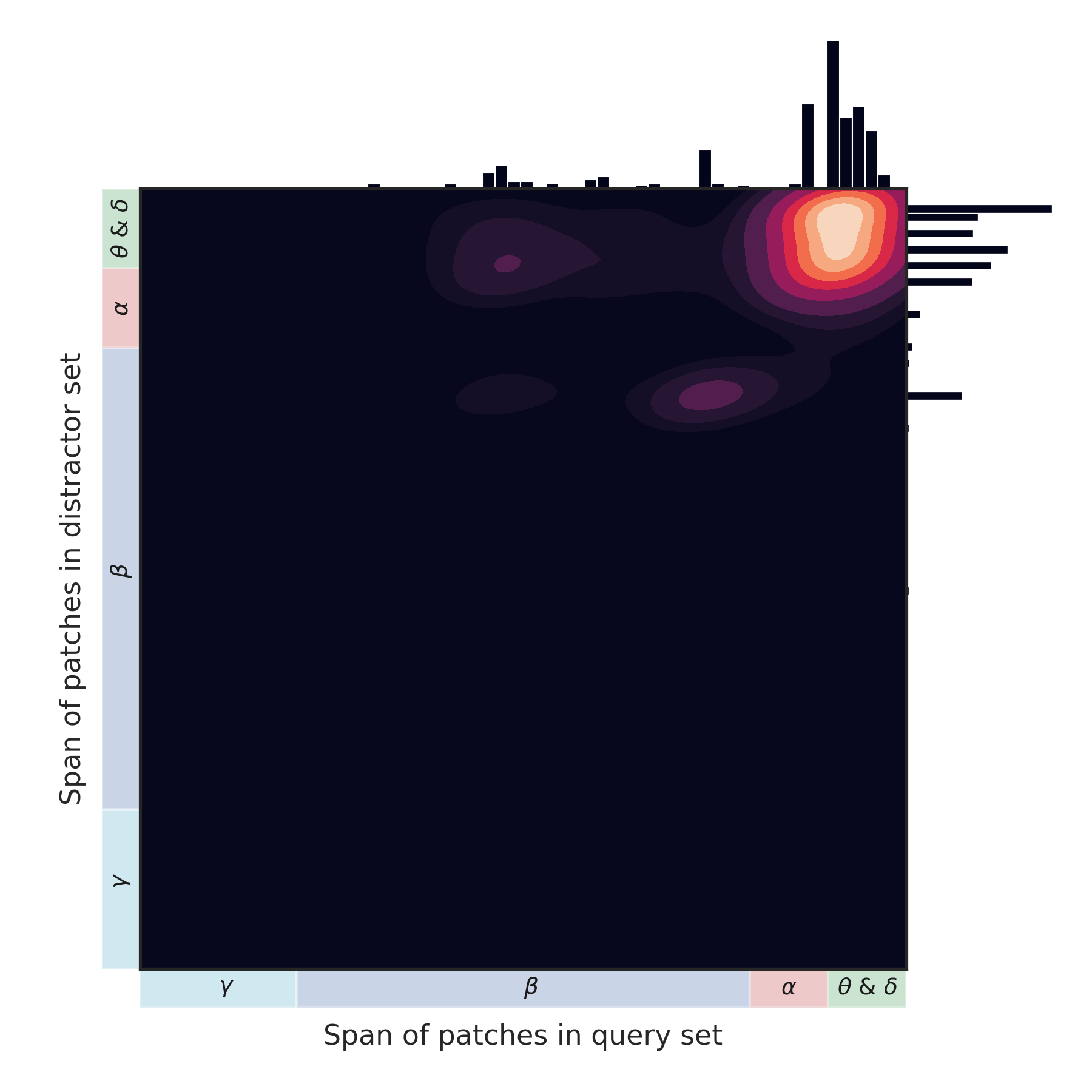}
    \end{minipage}
    \begin{minipage}{0.75\textwidth}
        \centering
        \includegraphics[width=0.32\linewidth]{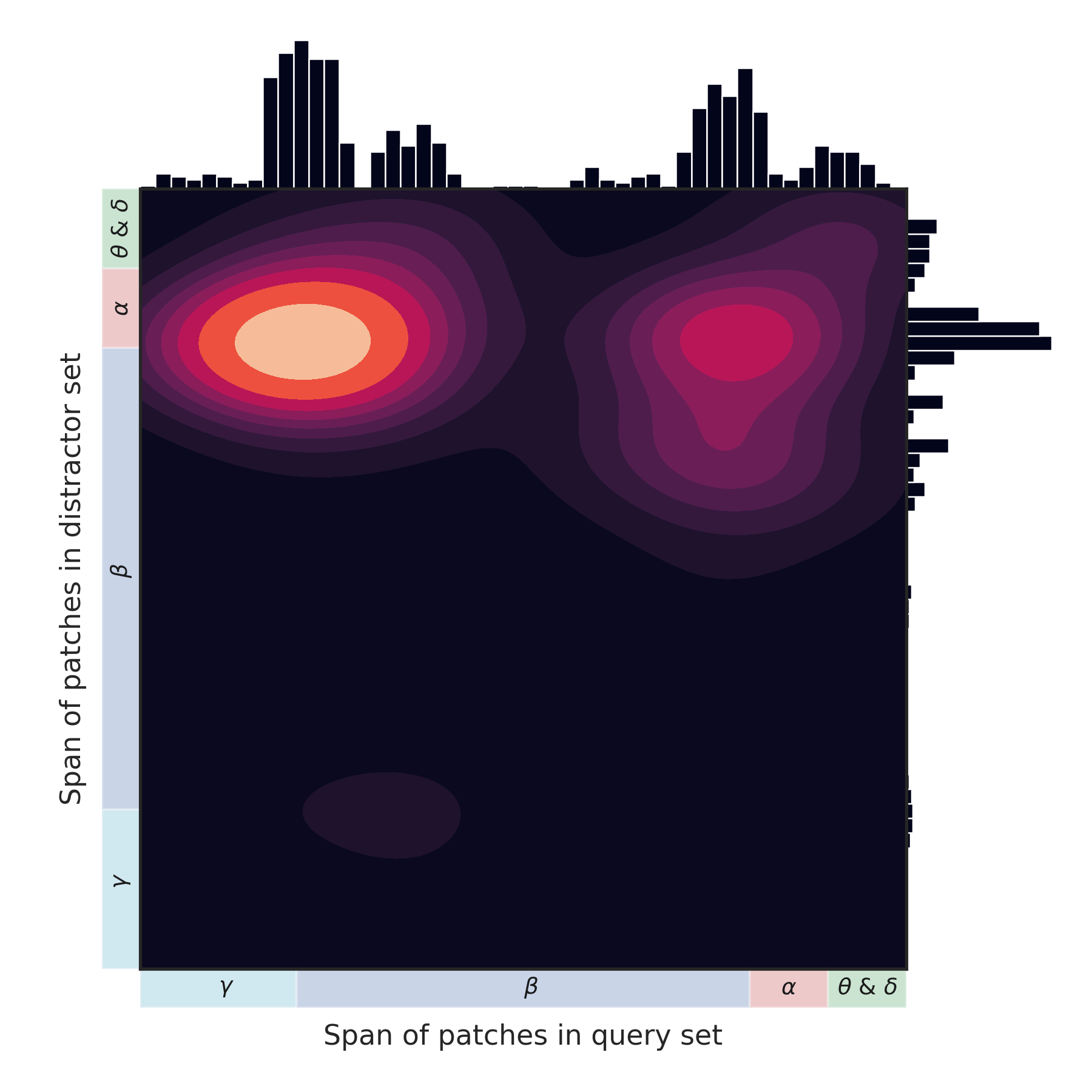}
        \includegraphics[width=0.32\linewidth]{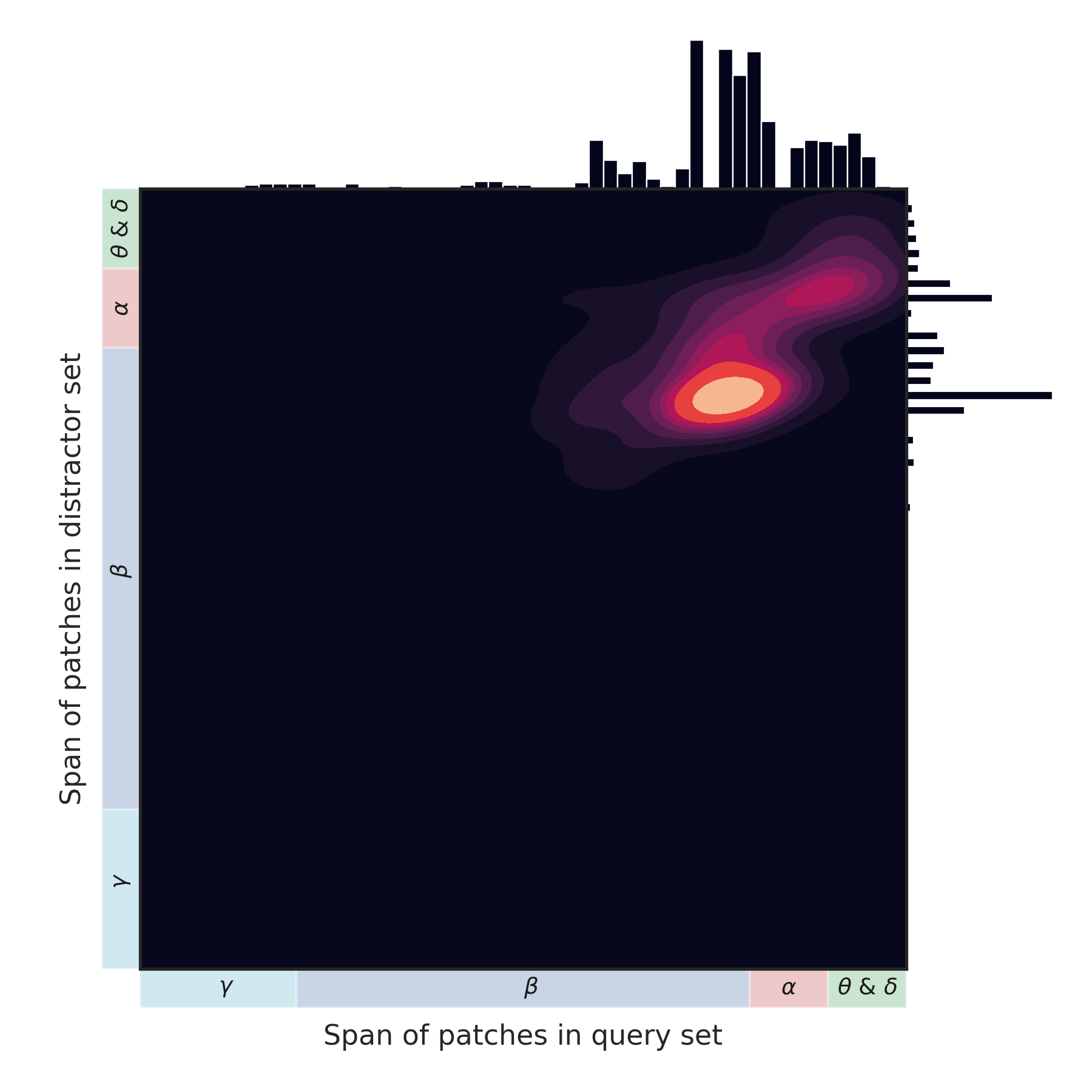}
        \includegraphics[width=0.32\linewidth]{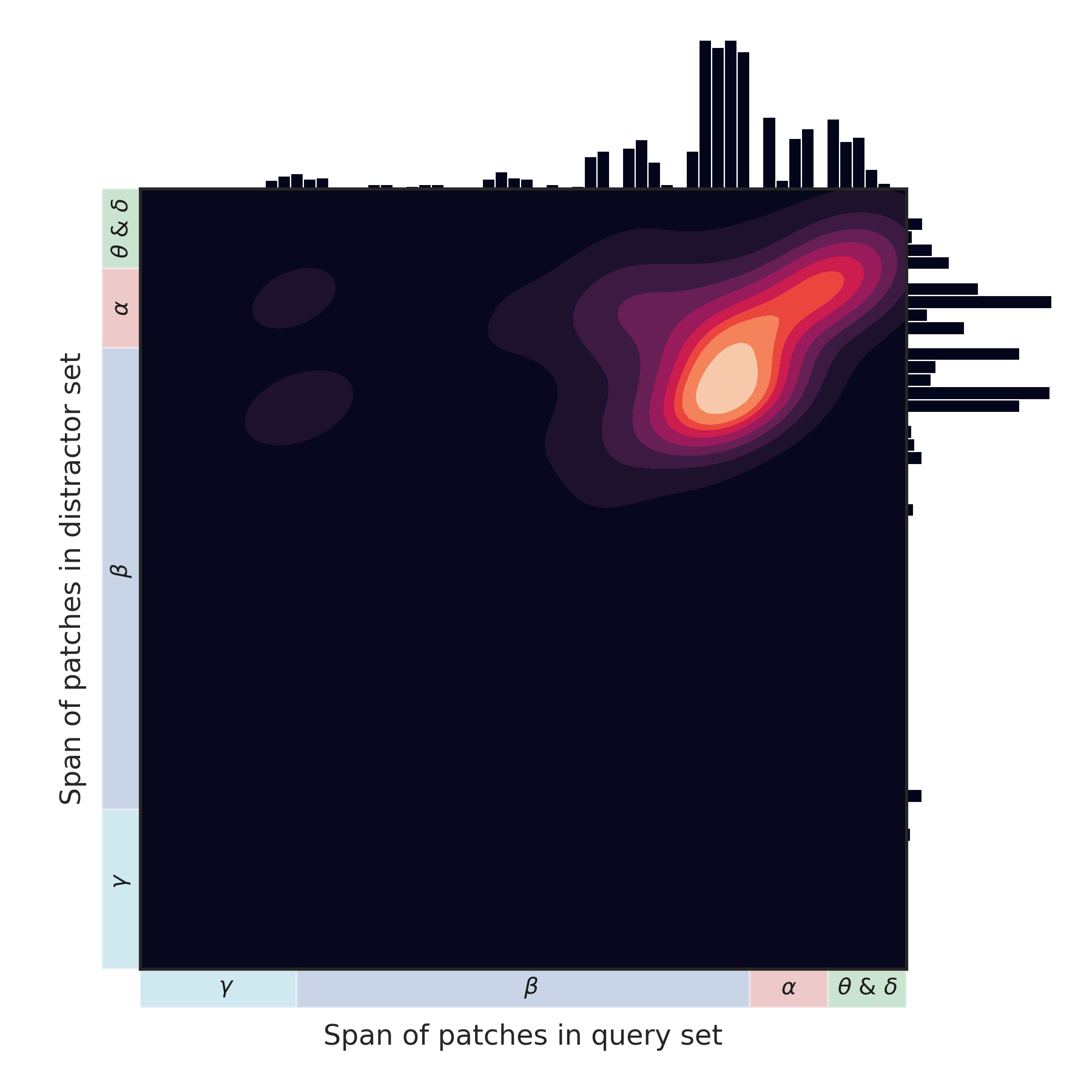}
    \end{minipage}
    \caption{Marginal histogram of KDE on edit pairs for 2-class and 18-class settings (row 1 and 2), respectively. (Left to right: underfitting, well-trained, and overfitting.)}
    \label{histogram}
\end{figure*}

In this section, we offer visual insights into counterfactuals using Fig.~\ref{vis_spec}, which is representative of the overall data. It displays the query and distractor pairs from a successful CE, with the left column representing the query and the right the distractor. Replacements are shown top to bottom, with two edits needed to change the model's prediction. Each replacement is marked by a red bounding box, predominantly located in the $\alpha$ band, corroborated by Table~\ref{results} and aligning with our assumptions. These replacements occur in distinct time periods, highlighting the model's capability to discern temporal differences. Interestingly, highlighted regions don't always correspond with the highest values, suggesting avenues for future research.
Furthermore, as depicted in Fig.~\ref{histogram}, we present a marginal histogram of kernel density estimation (KDE) for query and distractor spectrograms in 2-class and 18-class settings (first and second rows, respectively). We flatten the 7 $\times$ 7 square regions in Fig.~\ref{anno} (left), with each axis having a length of 49, representing query and distractor, and each 7 corresponds to a frequency range in Fig.~\ref{anno} (right). The histogram indicates the frequency of regions chosen, while the KDE shows the joint distribution of selected edit pairs. The lighter an area, the more often it is selected as the replacement region. We train the ResNet-18 classification model in three ways: underfitting, well-trained, and overfitting (left to right), aiming to observe variations in CE results. High Near-KP values suggest that highlighted areas should be near the intersection of $\alpha$ band, while high Same-KP values indicate they should align with the antidiagonal. In the underfitting case, the highlighted areas are dispersed across all frequencies for query spectrograms, indicating that the model has not learned distinctive classification features. Conversely, in the overfitting case, the highlighted areas gravitate towards the visually distinctive $\theta$ and $\delta$ bands, potentially biased by contradicting empirical priors. The well-trained cases show compact distributions in the $\alpha$ band, aligning with our expectations and suggesting that the model is likely making predictions based on truly meaningful features.

\section{CONCLUSION}
In this study, we introduce a CE method tailored for EEG MI classification. We have devised a comprehensive framework that adeptly adapts contemporary non-generative CE techniques and evaluation criteria to EEG data. Furthermore, we have developed specialized keypoints explicitly designed for EEG-based tasks. Our empirical results carry significant implications: they not only affirm the prowess of the pretrained model in modifying predictions with minimal edits, but they also resonate with established human empirical insights, reinforcing trust in the model's predictions.

This work lays a foundational pathway for researchers to delve into the realm of BCI counterfactual explanations, expanding beyond conventional explanatory tools such as SHAP and LRP. Our contribution delineates a streamlined yet robust procedure accompanied by a benchmark rooted in fundamental configurations. We posit that such novel explorations may substantially aid in elucidating intricate yet auspicious domains, such as decoding of imagined speech~\cite{imaginedspeech,b7,b8}. However, we acknowledge the scope for refinement in our approach. In future work, we aim to integrate various explanatory tools and also enhance the adaptability of the framework to accommodate diverse input dimensions. Furthermore, we intend to reassess and refine evaluation metrics to identify approaches most fitting for BCI applications.

\section*{Acknowledgement}
We thank Seung-Hyup Na and Yeong-Joon Ju for their valuable discussions that significantly contributed to the development of the research topic.

\bibliographystyle{IEEEtran}
\bibliography{bb}

\end{document}